\documentclass[aps,prl,twocolumn,showpacs]{revtex4}
\usepackage{dcolumn}
\usepackage{bm}
\usepackage{graphicx}
\usepackage{amsmath}
\usepackage{latexsym}
\usepackage{amsfonts}
\usepackage{amssymb}
\usepackage{array}
\usepackage{epsfig}

\newcommand{\ket}[1]{\left\vert#1\right\rangle}

\begin{document}
\title{Single-photon excitation of surface plasmon polaritons}

\author{M. S. Tame,$^1$ C. Lee,$^2$ J. Lee,$^{2,3}$ D. Ballester,$^1$ M. Paternostro,$^1$ A. V. Zayats,$^{4}$ and M. S. Kim$^1$}
\affiliation{$^1$School of Mathematics and Physics, Queen's University,~Belfast BT7 1NN, United Kingdom\\
$^2$Department of Physics, Hanyang University, Seoul 133-791, Korea \\
$^3$Quantum Photonic Science Research Center, Hanyang University, Seoul 133-791, Korea \\
$^4$Centre for Nanostructured Media, IRCEP, Queen's University, Belfast BT7 1NN, United Kingdom}

\date{\today}
   
\begin{abstract} 
We provide the quantum mechanical description of the excitation of surface plasmon polaritons on metal surfaces by single-photons. An attenuated-reflection setup is described for the quantum excitation process in which we find remarkably efficient photon-to-surface plasmon wavepacket-transfer. Using a fully quantized treatment of the fields, we introduce the Hamiltonian for their interaction and study the quantum statistics during transfer with and without losses in the metal. 
\end{abstract}

\maketitle

The emerging field of plasmonics~\cite{Zayats} is experiencing a considerable increase in interest from researchers in many areas of the physical sciences~\cite{photoncircuit}. 
Plasmonic-based nanophotonic devices in particular have begun to attract a keen interest from the quantum optics community
for their use in quantum information processing (QIP)~\cite{plasmonQIP,Alte,Lukin1,Lukin2}. In order to unlock the potential that plasmonics at the quantum level can offer, a clear understanding of the interplay between single-photons and surface plasmon polaritons (SPPs) is of fundamental importance. Recent studies have focused on various systems where SPPs and photons interact~\cite{Alte,Lukin1,Lukin2}.
However, a major obstacle has been the low transfer efficiencies found 
at the single-photon level~\cite{Alte, Lukin1}, thus a complete quantum description of an efficient transfer process
is highly desirable. 
With an extensive understanding of photon-SPP coupling in the quantum regime, we can expect to open up an array of new applications in QIP, based on linear and nonlinear plasmonic effects facilitated by strong electromagnetic field confinement~\cite{Lukin1,nonlinplasm}.

In this work we provide the first {\it quantum} description of the coupling between single photons and SPPs in a versatile attenuated-reflection (ATR) setup previously used only for {\it classical} SPP generation~\cite{Kret,Otto}. This is distinct from earlier work, such as couplings at rough surfaces~\cite{ER}, requiring an entirely different approach. 
The Hamiltonian that we introduce is based on a fully quantized treatment of both photon and SPP field modes and applies to a wide-range of ATR parameters. We find that remarkably high {\it quantum efficiencies} can be reached for photon-to-SPP transfer. We then establish the extent to which the excited SPPs preserve the quantum statistics of the photons as they travel on realistic metal surfaces. Our work provides significant insights into the physics of photon-SPP coupling at the quantum level. The methods developed are well-suited to other coupling geometries.

\begin{figure}[t]
\centerline{\psfig{figure=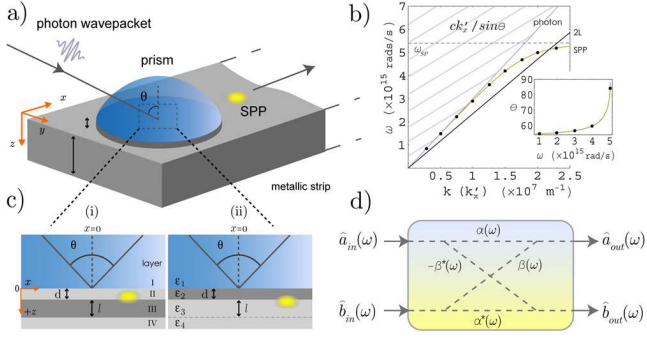,width=8.5cm}}
\caption{Single-photon excitation of SPPs using attenuated-reflection. 
{\bf (a)}: A photon wavepacket is injected into the system at a specific angle $\theta$, with a prism mediating an interaction between the photon and SPP modes. The minimum prism size is diffraction-limited.
{\bf (b)}: Dispersion relations for the photon (shaded region) and SPP (curve). The prism enables mode-matching. 
{\bf (c)}: Two ATR excitation geometries; 
(i) Otto and (ii) Kretschmann-Raether, see text for details. {\bf (d)}:~Transfer process for the photon and SPP mode operators.}
\label{fig1}
\end{figure}

SPPs are highly confined, nonradiative electromagnetic excitations associated with electron charge density waves propagating along a dielectric-metal interface. In Fig.~\ref{fig1}~{\bf (a)} we show the ATR setup utilized for single-photon excitation of SPPs. At various points we will introduce the metal as silver only to illustrate our main results; the theory developed here fits a far more general setting.  
For SPP excitations, due to the collective nature of the electron charge density waves, a macroscopic picture of the resulting electromagnetic field is appropriate~\cite{ER}. Upon quantization, SPPs therefore correspond to bosonic modes. The quantized vector potential in the continuum limit for SPPs propagating along an air-metal interface at 
$z=0$ in the $\hat{\mathbf x}$ direction, as shown on the right of Fig.~\ref{fig1}~{\bf (a)}, is given by~\cite{ER,Loudon} 
$\hat{\mathbf A}_{SPP}({\mathbf r},t)\propto \int_0^{\infty} {\mathrm d} \omega ({\cal N}(\omega)L)^{-1/2}[\phi({\mathbf r},\omega)e^{-i \omega t}\hat{b}(\omega)+ h.c]$. The dispersion relation is $\omega^2=c^2k^2(\epsilon_m+1)/\epsilon_m$ with $\epsilon_m$ the permittivity of the metal, ${\cal N}(\omega)$ is a frequency dependent normalization~\cite{ER} and $L$ is the {\it profile-width}~\cite{Loudon}. The $\hat{b}(\omega)$'s ($\hat{b}^{\dag}(\omega)$'s) correspond to annihilation (creation) operators which obey 
commutation relations $[\hat{b}(\omega),\hat{b}^{\dag}(\omega')]=\delta(\omega-\omega')$. The mode functions 
are given by $\phi({\mathbf r},\omega)=[(i \hat{\mathbf x}-k \hat{\mathbf z}/\nu)e^{-\nu z} \vartheta(z)+(i \hat{\mathbf x}+k \hat{\mathbf z}/\nu_0)e^{\nu_0 z} \vartheta(-z)]e^{i {\mathbf k} \cdot {\mathbf r}}$, where the wavevector ${\mathbf k}=k \hat{\mathbf x}$, $\vartheta(z)$ is the Heaviside step function and the decay of the SPP into the metal (air) is parameterized by $\nu^2=k^2-\epsilon_m \omega^2/c^2$ ($\nu_0^2=k^2-\omega^2/c^2$). 
For photons propagating in air 
in the $\hat{\mathbf k}'$ direction ($\hat{\mathbf k}'=\sin \theta \hat{\mathbf x}+\cos \theta \hat{\mathbf z}$, as shown on the left of Fig.~\ref{fig1}~{\bf (a)}), we have~\cite{Loudon} $\hat{\mathbf A}_{P}({\mathbf r},t)\propto \int_0^{\infty} {\mathrm d} \omega (\omega A)^{-1/2}[e^{i k'(\hat{\mathbf k}' \cdot {\mathbf r})} e^{-i \omega t}\hat{a}(\omega)+ h.c]$. The dispersion relation is $\omega=c k'$, $A$ is the beam cross-section and $[\hat{a}(\omega),\hat{a}^{\dag}(\omega')]=\delta(\omega-\omega')$.  
Here, the SPPs and photons are transverse magnetic modes. At the single-photon level only small intensities of the photon field are involved and any nonlinear terms in the photon-SPP coupling can be sufficiently neglected~\cite{linok}. We are thus led
to the following natural linear coupling Hamiltonian
for the entire system shown in Fig.~\ref{fig1}~({\bf a}),
\begin{eqnarray}
\hat{\cal H}_S&=& \int_{0}^{\infty}{\mathrm d} \omega \hbar \omega \hat{a}^{\dag}(\omega)\hat{a}(\omega)+\int_{0}^{\infty}{\mathrm d} \omega \hbar \omega \hat{b}^{\dag}(\omega)\hat{b}(\omega)  \label{Hamil} \\
& & + i \hbar \int_{0}^{\infty}{\mathrm d} \omega [g(\omega)\hat{a}^{\dag}(\omega)\hat{b}(\omega)-g^*(\omega)\hat{b}^{\dag}(\omega)\hat{a}(\omega)]. \nonumber
\end{eqnarray}
The first and second terms are the photon and SPP fields' free-energy respectively. 
The last term, which we denote as $\hat{\cal H}_{int}$, describes interactions between the two fields, where the coupling $g(\omega)$ is a function of the system parameters for a given ATR geometry.
In Fig.~\ref{fig1}~{\bf (b)} we show the dispersion relations for SPPs and photons. As an example, we choose $\epsilon_m=1-\omega_p^2/\omega^2+\delta \epsilon_m^r$, where $\omega_p=1.402 \times 10^{16}$rad/s is the plasma frequency for silver and $\delta \epsilon_m^r=29 \omega^2/\omega_p^2$ is a background correction term~\cite{JohnChrist}. Neglecting Ohmic losses, over the $\omega$ range of the SPP modes, $k$ produces a curve which 
approaches the surface plasma frequency $\omega_{sp}$, where $\epsilon_m=-1$. 
On the other hand, the $\hat{\mathbf x}$ component of ${\mathbf k}'$ for photons in air incident at angle $\theta$ 
covers the shaded region of Fig.~\ref{fig1}~{\bf (b)}. 
Two ATR geometries that can provide the necessary mode-matching for coupling photons to SPPs are shown in Fig.~\ref{fig1}~{\bf (c)}, denoted as (i)~Otto (O)~\cite{Otto} and (ii)~Kretschmann-Raether (KR)~\cite{Kret}. Both consist of a prism in layer I with permittivity $\epsilon_1$. The O (KR) geometry has air in layers II and IV (III and IV) with $\epsilon_2=\epsilon_4=1$ ($\epsilon_3=\epsilon_4=1$) and metal in layer III (II) with $\epsilon_3=\epsilon_m$ ($\epsilon_2=\epsilon_m$). In both, SPPs are excited on the II/III interface, with $z \to z-d$ in $\phi ({\mathbf r},\omega)$. For the KR geometry: $\nu \leftrightarrow \nu_0$.
The thickness $l$ is assumed to be far larger than the decay of the SPP into the metal (air), {\it i.e.} $l \gg \nu^{-1}$ ($\nu_0^{-1}$), making the effects of layer IV negligible.  

With the ATR setup introduced, we can now formulate the photon-SPP coupling model of Eq. (\ref{Hamil}).  
For each $\omega$ in both geometries the coupling can be described by a transfer matrix ${\cal T}(\omega)$ in the Heisenberg picture~\cite{salehleon} as
\begin{equation}
\left(
\begin{array}{c}
\hat{a}_{out}(\omega) \\
\hat{b}_{out}(\omega)
\end{array}
\right)=
\left(
\begin{array}{cc}
\alpha(\omega) & \beta(\omega) \\
-\beta^*(\omega) & \alpha^*(\omega)
\end{array}
\right)
\left(
\begin{array}{c}
\hat{a}_{in}(\omega) \\
\hat{b}_{in}(\omega)
\end{array}
\right).
\end{equation}
The transfer process is depicted in Fig.~\ref{fig1}~{\bf (d)}, where the commutation relations of the quantum operators $\hat{a}(\omega)$ and $\hat{b}(\omega)$ define the structure of ${\cal T}(\omega)$, while its coefficients ($|\alpha(\omega)|^2+|\beta(\omega)|^2=1, \forall \omega$) are determined from the overlap of system modefunctions. 
By solving Maxwell's equations across the first three layers shown in Fig.~\ref{fig1}~{\bf (c)}, 
one finds the modefunctions of the field in layers II and III: $\psi({\mathbf r}, \omega)=\{[(\varphi_1 e^{-\gamma_2 z}+\varphi_2 e^{\gamma_2 z})\hat{\mathbf x} + (\varphi_3 e^{-\gamma_2 z}+\varphi_4 e^{\gamma_2 z})\hat{\mathbf z}]\vartheta(z)\vartheta(d-z)+ [\varphi_5 e^{-\gamma_3 z} \hat{\mathbf x} + \varphi_6 e^{-\gamma_3 z} \hat{\mathbf z}]\vartheta(z-d) \}e^{i\kappa x}$. Here, the $\varphi_i$'s are constants related by boundary conditions at the interfaces,
$\gamma_i=(\kappa^2-\epsilon_i \omega^2/c^2)^{1/2}$ and the dispersion relation $\kappa=\sqrt{\epsilon_1}(\omega/c)\sin \theta$. The O (KR) geometry has $\epsilon_2=1$ ($\epsilon_m$) and $\epsilon_3=\epsilon_m$ ($1$) with $\epsilon_m=1-\omega_p^2/(\omega^2+i \omega \Gamma)+ \delta \epsilon_m$, which now includes a damping factor $\Gamma$ for the metal and a complex correction term $\delta \epsilon_m$~\cite{JohnChrist}. The complete modefunctions for the three-layer (3L) system are: $\Psi({\mathbf r}, \omega)=r \tilde{\psi}({\mathbf r}, \omega)\vartheta(-z)+ \tau \psi({\mathbf r}, \omega)\vartheta(z)$ where $r$ and $\tau$ ($|r|^2+|\tau|^2=1$) are obtained from Fresnel's relations at the boundaries. 
However, the $\tilde{\psi}({\mathbf r}, \omega)$ are not involved in the coupling due to mode-matching; they always have a real component of their wavevector in $\hat{\mathbf z}$. On the other hand, mode-matching can be satisfied between the two-layer (2L) modefunctions $\psi({\mathbf r}, \omega)$ and the SPP modefunctions by fixing the angle $\theta$ correctly. 
For instance, by setting $\kappa=k$ the dispersion lines cross at $\theta=\sin^{-1}[\epsilon_m/(\epsilon_1(1+\epsilon_m))]^{1/2}$ in both geometries. In Fig.~\ref{fig1}~{\bf (b)} we show this for a particular angle $\theta=85^\circ$ (2L line). The inset shows that mode-matching over the entire range of $\omega$ can be achieved, for example, using a prism with $\epsilon_1=1.51$ and silver with $\Gamma=6.25 \times 10^{13}$rad/s and $\delta \epsilon_m=\delta \epsilon_m^r+i \delta \epsilon_m^i$, where 
$\delta \epsilon_m^i=0.22$~\cite{JohnChrist}. This range is important for excitation with a photon wavepacket of finite width, as we show later, and is not possible in other excitation schemes such as the grating-type coupler.

\begin{figure}[t]
\centerline{
\psfig{figure=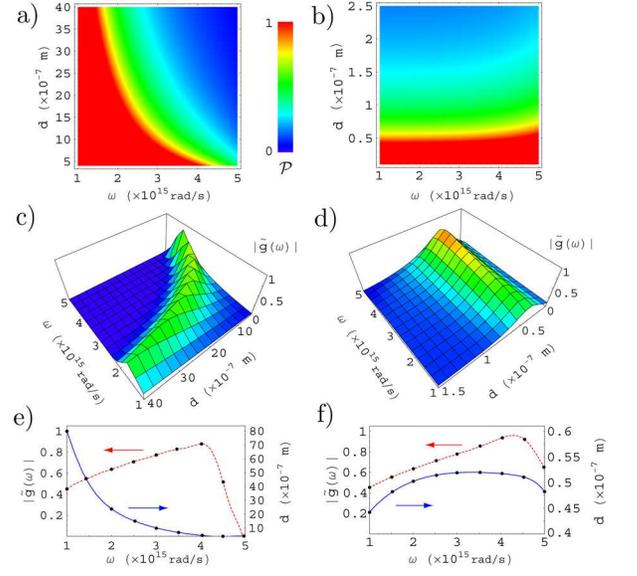,width=7.9cm}}
\caption{Photon-SPP coupling: The left (right) column corresponds to the O (KR) geometry. Panels {\bf(a)} and {\bf(b)} depict the penetration-factor ${\cal P}$. 
Panels {\bf(c)} and {\bf(d)} show the behavior of the coupling $|\tilde{g}(\omega)|=\frac{2}{\pi}|g(\omega)|$. The optimal values of $|\tilde{g}(\omega)|$ (dashed line) are displayed in panels {\bf(e)} and {\bf(f)} along with the values of $d$ (solid line) at which they occur.}
\label{fig2}
\end{figure}

In Fig.~\ref{fig1}~{\bf (d)} the $\hat{b}_{in/out}(\omega)$ operators are associated with the in/out SPP modefunctions $\phi({\mathbf r}, \omega)$ ($\hat{b}_{in}(\omega)=\hat{b}(\omega)$) and $\hat{a}_{in/out}(\omega)$ with the in/out 3L modefunctions $\Psi({\mathbf r}, \omega)$. For negligible loss on entry into the prism medium, we can assume the operator relation $\hat{a}_{in}(\omega)=\hat{a}(\omega)$. 
We then have $\beta^*(\omega)=-\tau[\delta(\omega-\omega')\delta(k-\kappa)\int{\rm d}z ({\cal N}_1^{-1/2}(\omega)\phi({\mathbf r}, \omega))^*\cdot ({\cal N}_2^{-1/2}(\omega')\psi({\mathbf r}, \omega'))]$~\cite{SPPparams}. 
Several factors permit the use of the mode overlap in the value of $\beta^*(\omega)$. First, we assume that the SPP modes experience negligible damping during the excitation process, ${\rm Im}(\epsilon_m) \approx 0$, imposing damping effects subsequently as the SPP propagates. 
Second, the SPP is assumed to exit the prism region on a time-scale such that mode-matching conditions are broken 
almost 
immediately after excitation. This can be achieved by adjusting the excitation point~\cite{RSPP}.
Third, as the SPP modefunctions exist in the region $z \in (-\infty,\infty)$,
$d$ must be chosen such that their decay into the prism is negligible, allowing it
to be neglected from their definitions.
To check an acceptable range of $d$ we define a {\it penetration-factor} ${\cal P}=2/\nu_0 d$ ($2/\nu d$) for the O (KR) geometry and consider the SPP modes 
as good approximations for ${\cal P}\leq 1$, where $|\phi({\mathbf r}, \omega)|^2$ at $z=0$ is less than $2\%$ its maximum value. In Figs.~\ref{fig2}~{\bf (a)} and {\bf (b)} we use the example of silver to show ${\cal P}$ over a range of $\omega$ and $d$ for the  
two ATR geometries.

With the above considerations, we can now determine the coupling $g(\omega)$.
In order to connect $\beta^*(\omega)$ and $g(\omega)$, 
we set $\alpha=\cos \Theta$ and $\beta=e^{i\Phi}\sin \Theta$ ($\Theta \in [0,\pi/2]$, $\Phi \in [0,2\pi]$) for each $\omega$ and parameter set $\{ d, \epsilon_i, \theta \}$ for a given geometry. The corresponding Hamiltonian in the Schr\"odinger picture is $\hat{\cal H}_{int}$ from Eq.~(\ref{Hamil}) with $g(\omega)=e^{i \Phi(\omega)}\Theta(\omega)$~\cite{salehleon}. 
In Figs.~\ref{fig2}~{\bf (c)} and {\bf (d)} we use silver to plot a rescaled coupling, $|\tilde{g}(\omega)|=\frac{2}{\pi}|g(\omega)|$, for the two ATR 
geometries, 
such that $|\tilde{g}(\omega)|= 1$ (0) corresponds to a unit (zero) transfer probability of a photon to an SPP. Figs.~\ref{fig2}~{\bf (a)} and {\bf (b)} show the optimal $|\tilde{g}(\omega)|$ for both geometries 
satisfy ${\cal P} \lesssim 1$. In Figs.~\ref{fig2}~{\bf (e)} and {\bf (f)} we plot these optimal values and the value of $d$ at which they occur. 
The optimal $|\tilde{g}(\omega)|$ in both geometries rises for increasing $\omega$, reaching an apex, then drops sharply as $\omega$ tends toward 
$\omega_{sp}$. This behavior is due to a dominance of the value for $\tau$ in $\beta^*(\omega)$ 
at large $\omega$, which decreases rapidly due to boundary conditions and the large $\theta$ required for mode-matching. Such excellent coupling values have been found {\it classically}~\cite{Otto,Kret}, however, this is the first time a rigorous {\it quantum mechanical} treatment has been achieved, making it possible for us to determine correctly the quantum efficiency of single-photon excitation. The coupling $g(\omega)$ cannot be deduced from a classical model of the system.

So far we have focused on single modes of the system. However, it is important to consider the transfer of a photon wavepacket, such as in an experiment, to an SPP wavepacket state. An $n$-photon wavepacket state is given by $\ket{n_{\xi}}=(n!)^{-1/2}(\hat{a}_{\xi}^{\dag})^n \ket{0}$, where $\hat{a}_{\xi}^{\dag}=\int {\rm d} \omega \xi(\omega)\hat{a}^{\dag}(\omega)$ with $\int {\rm d} \omega |\xi(\omega)|^2=1$~\cite{Loudon}. For simplicity we take a Gaussian profile $\xi(\omega)$ 
for a wavepacket produced at time $t_0=0$ with bandwidth $\Delta \omega=2\sigma \sqrt{2 \log 2}$ and central frequency $\omega_0$. 
We allow $\omega \in (-\infty,\infty)$ as $\Delta \omega \ll \omega$.
For wavepacket-transfer with negligible {\it deformation}, each $\omega$ must have approximately the same $g(\omega)$ and $\theta$. This is satisfied given a small enough bandwidth with slowly varying $g(\omega)$ and $\theta$. For example, using silver with $\Delta \lambda=10$nm, one finds that $\Delta \theta$ rises exponentially 
from $0.004^\circ$ at $1 \times 10^{15}$rad/s to $14.61^\circ$ at $5 \times 10^{15}$rad/s, which can be attributed to the dependence of $\theta$ on $\omega$ (see inset of Fig.~\ref{fig1}~{\bf (b)}). The couplings show a similar behavior, with $|\Delta g(\omega)|=0.01$ for $d$ optimizing $|g(\omega_0)|$ at $1 \times 10^{15}$rad/s, rising to $|\Delta g(\omega)|=0.2$ ($0.04$) at $5 \times 10^{15}$rad/s for the O (KR) geometry. The large difference is due to the sharper drop in $|g(\omega)|$ for O 
at high $\omega$.
Significant wavepacket deformation can 
be avoided by operating at low $\omega$, although at the expense of the coupling. In general, a narrow bandwidth will provide access to larger couplings with negligible deformation.

\begin{figure}[t]
\centerline{
\psfig{figure=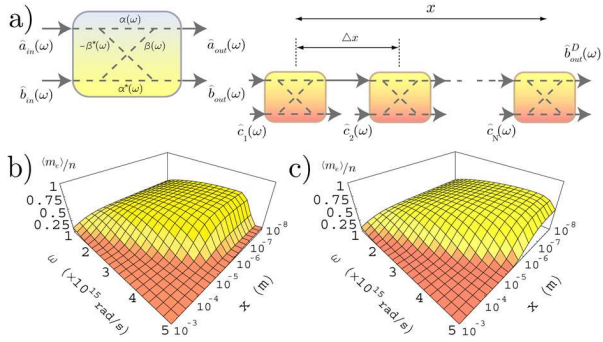,width=7.9cm}}
\caption{Damping model for propagating SPPs. {\bf (a)}: Bath modes interact with the SPP mode as it travels along the metal. Panels {\bf (b)} and {\bf (c)} show the normalized expected mean SPP-count $\langle m_e \rangle/n$ at point $x$ for an $n$ photon-to-SPP wavepacket transfer via the O and KR geometries respectively.}
\label{fig3}
\end{figure}

Finally we turn our attention to damping in the metal as the excited SPP propagates. For the coupling, the approximation ${\rm Im}(\epsilon_m)\approx 0$ was made for the excited SPP. However as it travels, finite conductivity of the metal and surface roughness result in heating and radiative losses respectively~\cite{Zayats}; for a reasonably smooth surface, thermal loss is the main source of damping.
While a quantization of 
the decayed SPP modes can be 
performed, a mathematically equivalent and simpler model is the method of arrays shown in Fig.~\ref{fig3}~{\bf (a)}~\cite{Loudon}. Here 
we introduce 
a bath of field modes, described by operators $\hat{c}_i(\omega)$ ($i=1,.., N$) separated by $\Delta x$,
interacting with the SPP wavepacket as it propagates. 
In the limit $N \to \infty$ and $\Delta x \to 0$, the SPP operator becomes $\hat{b}_{out}^D(\omega)=e^{iKx}\hat{b}_{out}(\omega)+i\sqrt{2 \kappa(\omega)}\int_{0}^{x}{\rm d} x' e^{iK(x-x')}\hat{c}(\omega,x')$,
with 
$\hat{c}_i(\omega) \to \sqrt{\Delta x} \hat{c}(\omega,x')$ and $\delta_{ij} \to \Delta x \delta(x'-x'')$. The array coefficients are chosen such that the bath modes induce a change in the SPP wavevector $k$ matching that of the complex $\epsilon_m$, {\it i.e.} $k \to K=(\omega/c)[\epsilon_m/(1+\epsilon_m)]^{1/2}=k+i\kappa(\omega)$, where $2\kappa(\omega)$ is the loss per unit length of propagation. 
We then set the relations $\langle \hat{c}(\omega,x') \rangle=\langle \hat{c}^{\dag}(\omega,x') \rangle=\langle \hat{c}^{\dag}(\omega,x')\hat{c}(\omega,x'') \rangle=0$ 
for the bath modes at room temperature and the frequencies considered~\cite{Loudon}. 
We assume that the excited SPP wavepacket with $\omega_0$ has a narrow enough bandwidth such that $\kappa(\omega)\approx\kappa(\omega_0)=\kappa_0$ and $k\approx k(\omega_0)+(\omega-\omega_0)v_G^{-1}(\omega_0)$, with $v_G^{-1}(\omega_0)=\frac{\partial k (\omega)}{\partial \omega}|_{\omega=\omega_0}$. The flux of SPPs at point $x$ along the metal surface 
is then simply $f_{out}(t)=\langle \hat{b}_{out}^{D \dag}(t)\hat{b}_{out}^D(t) \rangle=e^{-2\kappa_0 x} \langle \hat{b}_{out}^{\dag}(t_R)\hat{b}_{out}(t_R) \rangle$, where $t_R=t-x v_G^{-1}(\omega_0)$. 
For an initial SPP wavepacket with $n$ excitations, $\langle \hat{b}_{out}^{\dag}(t_R)\hat{b}_{out}(t_R) \rangle=n |\tilde{\xi} (t_R)|^2$. A detector with efficiency $\mu$~\cite{Loudon,LoudonLoss} operating for time period $[x v_G^{-1}(\omega_0)-1/\sigma, x v_G^{-1}(\omega_0)+1/\sigma]$ would measure a mean SPP-count of $\langle m \rangle=\mu \int{\rm d} t f_{out}(t)=\mu n e^{-2\kappa_0 x}$. The Photon-to-SPP transfer process must however be incorporated for determining the {\it expected} mean SPP-count $\langle m_{e}\rangle$ from an incident $n$ photon wavepacket. The entire process is analogous to an inefficient detection problem~\cite{LoudonLoss} and 
we have $\langle m_e \rangle=\mu |\beta(\omega_0)|^2 n e^{-2 \kappa_0 x}$. In Figs.~\ref{fig3}~{\bf (b)} and {\bf (c)} we show $\langle m_{e} \rangle/n$ 
for 
the ATR geometries, where $\mu=0.65$ is chosen as an example of non-ideal signal extraction using, {\it e.g.} a prism and photodetector. The detection of many identical excitations from a set rate of single photons would be required to determine $\langle m_{e}\rangle$. 

While the quantum observable $\langle m_e \rangle$ matches well the behavior of its classical counterpart, the field intensity $I$~\cite{Zayats}, it is not sufficient in an experiment to show that the SPPs are quantum excitations. We now consider another observable, the 
zero time-delay second-order quantum coherence function $g^{(2)}(0)$~\cite{Loudon} at a fixed position, defined as $g^{(2)}(0)\!=\!\langle\,\colon\!\hat{I}^{2}(t)\colon\!\rangle/\langle\,\colon\!\hat{I}(t)\colon\!\rangle^2$. Here, $\hat{I}$ is the intensity of the quantized field operator, $\colon\colon$ denotes normal-ordering and the expectation value is taken over the {\it initial} state of the field. 
For a classical field   
$1 \le g^{(2)}(0) \le \infty$. On the other hand, for an incident $n$ photon wavepacket $g^{(2)}(0)=\langle m(m-1) \rangle/\langle m \rangle^{2}$, where $\langle m \rangle=n\int_t^{t+T}{\rm d}t'|\tilde{\xi}(t')|^2$ and $\langle m(m-1) \rangle=n(n-1)[\int_t^{t+T}{\rm d}t'|\tilde{\xi}(t')|^2]^2$, giving $g^{(2)}(0)=1-1/n$. This always lies in the classically {\it forbidden} region $g^{(2)}(0)<1$. The value of $g^{(2)}(0)$ for an excited SPP wavepacket at point $x$ 
can be found by recognizing that the photon-to-SPP transfer and SPP propagation stages constitute an array of lossy beamsplitters~\cite{LoudonLoss}. At a beamsplitter with loss coefficient $\eta^{1/2}$, the quantum observables $\langle m\rangle \to \eta \langle m\rangle$ and $\langle m(m-1) \rangle \to \eta^2 \langle m(m-1) \rangle$. Thus, the individual losses accumulated cancel, leaving $g^{(2)}(0)$ surprisingly unaffected. A Hanbury-Brown Twiss type experiment~\cite{HBT} 
could be used to measure $g^{(2)}(0)$. 

We have provided the first quantum description of the Photon-to-SPP transfer process for ATR excitation. 
Remarkably good quantum couplings over a wide-range of frequencies were found.  
We also examined the extent to which the excited SPPs preserve quantum statistical properties. 
The techniques developed here provide key insights into the formulation of quantum descriptions for the photonic excitation of SPPs. This work can therefore be seen as an important starting point for future research into the design of new quantum plasmonic devices for applications based at the nanoscale, such as SPP-enhanced nonlinear photon interactions and SPP-assisted photonic quantum networking and processing.

{\it Acknowledgments}.- 
We thank M. D. Lukin, S. Song and S. K. Ozdemir for helpful discussions and acknowledge funding from EPSRC, QIPIRC, KRF~(2005-041-C00197) and ESF.


\begin{thebibliography}{99}
\bibitem{SPPparams} The normalizations ${\cal N}_i$ are carried out over $z \in(- \infty, \infty)$.
\end{thebibliography}
\end{document}